\pdfoutput=1

\documentclass[11pt]{article}

\usepackage[preprint]{acl}

\usepackage{times}
\usepackage{latexsym}

\usepackage[T1]{fontenc}

\usepackage[utf8]{inputenc}

\usepackage{microtype}

\usepackage{inconsolata}

\usepackage{graphicx}
\usepackage{xspace}
\usepackage{caption}
\usepackage{subcaption}
\usepackage{amsmath}
\usepackage{multirow}

\title{Does Few-Shot Learning Help LLM Performance in Code Synthesis?}

\author{
Derek Xu\textsuperscript{1}\textsuperscript{*}, Tong Xie\textsuperscript{1}\textsuperscript{*}, Botao Xia\textsuperscript{1}\textsuperscript{*}, Haoyu Li\textsuperscript{2}\textsuperscript{*} \\\AND
Yunsheng Bai\textsuperscript{3}, Yizhou Sun\textsuperscript{1}, Wei Wang\textsuperscript{1} \\\AND
\\
 \textsuperscript{1}University of California Los Angeles \\
 \textsuperscript{2}University of Illinois Urbana-Champaign \\
 \textsuperscript{3}Nvidia
\\
\small{\textsuperscript{*} Equal Contribution}
\\
}

\newcommand{\oursfree}{\textsc{CodeExemplar-Free}\xspace}
\newcommand{\oursmodel}{\textsc{CodeExamplar-Base}\xspace}
\newcommand{\perptgt}{\textsc{Perplexity (Target)}\xspace}
\newcommand{\perpsrc}{\textsc{Perplexity (Source)}\xspace}
\newcommand{\llama}{\textsc{Llama}\xspace}
\newcommand{\codellama}{\textsc{CodeLlama}\xspace}
\newcommand{\mixtral}{\textsc{Mistral}\xspace}
\newcommand{\tfsmall}{\textsc{T5-Small}\xspace}
\newcommand{\tfbase}{\textsc{T5-Base}\xspace}
\newcommand{\hevp}{\textsc{HumanEval+}\xspace}

\begin{document}
\maketitle

\begin{abstract}
Large language models (LLMs) have made significant strides at code generation through improved model design, training, and chain-of-thought. However, prompt-level optimizations remain an important yet under-explored aspect of LLMs for coding. This work focuses on the few-shot examples present in most code generation prompts, offering a systematic study on whether few-shot examples improve LLM's coding capabilities, which few-shot examples have the largest impact, and how to select impactful examples. Our work offers 2 approaches for selecting few-shot examples, a model-free method, \oursfree, and a model-based method, \oursmodel. The 2 methods offer a trade-off between improved performance and reliance on training data and interpretability. Both methods significantly improve \codellama's coding ability across the popular \hevp coding benchmark. In summary, our work provides valuable insights into how to pick few-shot examples in code generation prompts to improve LLM code generation capabilities. 
\end{abstract}

\section{Introduction}

Recently, large language models (LLMs)~\citep{radford2019language} have demonstrated impressive capabilities outside of natural language processing, including in mathematics~\citep{touvron2023llama,luo2023wizardmath}, time series forecasting~\citep{ansari2024chronos,woo2024unified,das2023decoder}, tabular data understanding~\citep{hegselmann2023tabllm,hollmann2022tabpfn,xu2024mixture}, and multi-modal understanding~\citep{liu2024visual}. Among these capabilities, LLMs' application to software engineering is particularly exciting. In just a few months, LLMs exhibit zero-shot abilities in code completion~\citep{peng2023impact}, code generation~\citep{roziere2023code,guo2024deepseek,dubey2024llama}, test case generation~\citep{vikram2023can,lemieux2023codamosa,schafer2023empirical}, debugger interaction~\citep{islam2024mapcoder,ho2024verilogcoder}, and repository-level generation~\citep{shrivastava2023repository,bui2024rambo,wang2024repogenreflex}.

In this work, we study the important task of code generation~\citep{roziere2023code}, where an LLM agent generates code described by a prompt consisting of a natural language function description and few-shot input-output examples. Current LLMs exhibit code generation capabilities through improved model design~\citep{guo2024deepseek}, training~\citep{roziere2023code}, and chain-of-thought~\citep{li2023structured}. Our work aims to enhance existing LLMs for code generation by improving the prompt itself, which remains an under-explored research problem. In fact, existing techniques evaluate on predefined prompt templates with little to no modifications~\citep{austin2021program,liu2024your}. 

To improve the prompt itself, we break down the prompt template into 2 components: (1) a natural language description, providing a high-level description of the code, and (2) few-shot input-output examples, describing function input outputs to disambiguate the natural language description. For example, the description ``return a list with elements incremented by 1'' can be disambiguated by the example ``incr\_list([1,2,3]) == [2,3,4],'' which shows that each numerical element of the list should be incremented by 1. 


Inspired by existing work~\citep{liu2021makes, liu2024large} that show LLM's in-context learning (ICL) ability is shaped by which examples are included in the ICL prompt, we hypothesize LLM's coding ability is also shaped by which few-shot examples are included in the code generation prompt~\footnote{Unlike in-context learning (ICL), few-shot examples in code synthesis prompts are not in the same input output space. Hence, existing ICL techniques for example selection~\citep{liu2021makes} does not work in our setting.}. We confirm our hypothesis on several language models: \tfsmall, \tfbase, \mixtral, and \codellama on the \hevp benchmark. Furthermore, we provide analysis on which examples contribute most to LLM's coding capabilities.


Given few-shot examples greatly affect LLM coding capability, we provide two methods for selecting few-shot examples: (1) a model-free algorithm, \oursfree, that picks examples based on an input metric, and (2) a model-based algorithm, \oursmodel that picks examples based on a bootstrapped training dataset. The former offers an interpretable data-agnostic algorithm, and the latter offers a better performing data-driven model. Both approaches support arbitrary token cost constraints. Both approaches substantially improve \codellama's coding capabilities on the \hevp benchmark under fixed token constraints. We summarize our contributions:

\begin{itemize}

    \item We demonstrate that choice of few-shot examples in the LLM prompt has a significant effect on LLM coding capabilities across 5 different LLMs. 
    \item We propose an interpretable model-free algorithm, requiring no training, that improves LLM code generation ability by only modifying the input prompt.
    \item We propose a data-driven neural network, trained on a dataset of code generation prompts, that improves LLM code generation by only modifying the input prompt. Both algorithms are gray-box: they do not require access to ground-truth weights of the model, only the logits for given input.

\end{itemize}

\begin{figure*}
    \centering
    \includegraphics[width=1.04\linewidth]{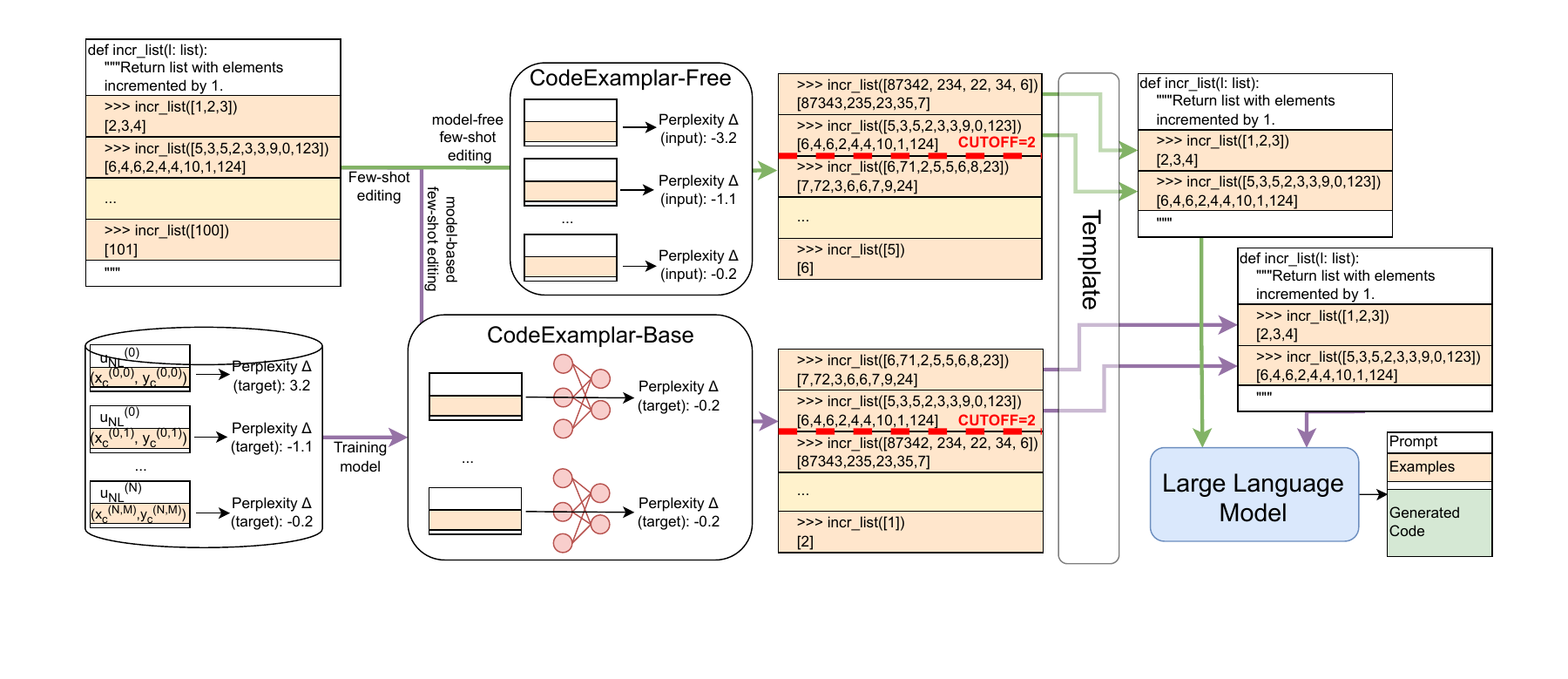}
    \vspace{-1.7cm}
    \caption{Overview or MetaNet Ranker and Perplexity Ranker. We use a ranking algorithm, $f_p$, to select the top $N=2$ examples to form a prompt that is fed to the LLM testing its coding capabilities.}
    \label{fig:main}
\end{figure*}

\section{Related Work}

\subsection{Code LLMs}

Recent advancements in large language models (LLMs) have led to significant progress in code generation and understanding. State-of-the-art models like GPT-4~\cite{openai2023gpt4}, Claude 3.5~\cite{claude3.5}, Llama 3~\cite{dubey2024llama}, and Mixtral~\cite{mixtral_8x22b_instruct_v0_1} have demonstrated remarkable code synthesis capabilities across various programming languages. These models leverage vast amounts of code data and natural language descriptions to generate contextually relevant and syntactically correct code snippets. More specialized code LLMs, such as CodeT5~\cite{wang2021codet5}, InCoder~\cite{fried2022incoder}, DeepSeek-Coder~\cite{guo2024deepseek}, StarCoder2~\cite{lozhkov2024starcoder}, Qwen2.5-coder~\cite{hui2024qwen2}, Magicoder~\cite{magicoder}, Artigenz-Coder~\cite{artigenz2024artigenz}
 etc. have been fine-tuned specifically for programming tasks, showing improved performance in code completion and bug fixing. Recent work has also explored the use of LLM-powered agents for more complex coding tasks. For instance, AgentCoder~\cite{huang2023agentcoder} and OpenCodeInterpreter~\cite{zheng2024opencodeinterpreter} introduce frameworks that decompose coding problems into subtasks or leverage execution feedback and iterative refinement, tackling challenging programming problems more effectively.

\subsection{In-Context Learning}
In-context learning enables LLMs to adapt to new tasks by providing examples within the input prompt \cite{brown2020language}.  Research focuses on optimizing example selection and ordering  \cite{lu2021fantastically,liu2021makes,wang2024large,peng2024revisiting}, including using curriculum learning~\cite{liu2024let} and comparing it with fine-tuning \cite{mosbach2023few}. This has driven advances in example construction \cite{chan2022data}, mechanism exploration \cite{rubin2021learning}, and reliability improvement \cite{gao2024customizing}.  However, challenges persist in performance stability and prompt sensitivity \cite{zhao2021calibrate}, motivating research on meta-learning \cite{coda2023meta} and model scale effects~\cite{sun2024generative}. To enhance inductive reasoning, researchers are developing techniques like Case2Code \cite{shao2024case2code} and SolverLearner \cite{cheng2024inductive} to evaluate and improve LLMs' ability to learn from input-output examples. These efforts, along with frameworks like Modelizer \cite{mammadov2024learning} and AcTracer \cite{huang2024active}, underscore the importance of selecting informative input-output pairs for strong reasoning and generalization.

\section{Problem Setting}


Our work studies how to generate effective prompts for large language model (LLM) code generation. We denote the textual prompt as $x_m^{(j)} \in \mathcal{X}_m$ and corresponding code (in text form) as $y_m^{(j)} \in \mathcal{Y}_m $. Each prompt consists of a natural language description, $u_{NL}^{(j)}\in \mathcal{U}_{NL}$, and $N$ few-shot demonstration examples, $D_j=[(x_c^{(j,i)}, y_c^{(j,i)})]_{i=0}^{N}$, where  $y_c^{(j,i)} \in \mathcal{Y}_c$ is the expected output object when the code is run on example input object $x_c^{(j,i)} \in \mathcal{X}_c$. A template, $\mathcal{T}$, converts the natural language description and few shot examples into said textual prompt: $x_M^{(j)} = \mathcal{T}(u_{NL}^{(j)},D_j)$. The LLM, $f_{m}:\mathcal{X}_m \rightarrow \mathcal{Y}_m$, generates code (in text form) provided the input prompt. To evaluate the LLM's coding ability, we check the Pass@1 metric~\citep{roziere2023code} of the code (executable), $f_c: \mathcal{X}_c \rightarrow \mathcal{Y}_c$ generated from the code (in text form) using a compiler, $\mathcal{C}$, as follows: $f_c^{(j)} = \mathcal{C}(x_m^{(j)}, y_m^{(j)})$.

Our work focuses on the different input output pairs, $D_j$, included in the prompt, $x_M^{(j)}$. Specifically, given a larger pool of $M\gg N$ demonstration examples, $\hat{D}_j = [(x_c^{(j,i)}, y_c^{(j,i)})]_{i=0}^{M}$, we propose a sorting algorithm that ranks examples by how much they would improve the Pass@1 score, $f_{p}: \mathcal{U}_{NL} \times (\mathcal{X}_c \times \mathcal{Y}_c)^M \rightarrow S_m$, where $S_n$ denotes the space of indices i.e. permutations of $\{0, 1, ..., M-1\}$. Our algorithm construct new prompts by selecting the top $N$ example prompts: $\tilde{D}_j = [(x_c^{(j,i)}, y_c^{(j,i)})]_{i\in f_p(D_j)[:N]}$\footnote{Note, our model design assumes each example is added independently, where the $(K+1)$th best example is not dependent on which example is the $K$th best. The independence assumption allows us to efficiently select an aribtrary cutoff $N$. We leave dependent example selection to future work.}. Directly including all $M$ examples from the larger pool, $\hat{D}_j$, into the prompt is infeasible, due to LLM's high compute costs on very long context prompts. In practice, large pools of test cases, $\hat{D}_j$, can either originate from human generated documentation~\cite{nassif2021generating,lemieux2023codamosa} or GPT-generated unit tests~\citep{austin2021program}. 

Compared to existing work which uses few human defined examples, $D_j$, our work algorithmically filters the best $N$ few-shot examples, $\tilde{D}_j$, from a larger pool of $M$ examples, $\hat{D}_j$, through a sorting algorithm, $f_p$\footnote{We propose both a model-free and a model-based version of the sorting algorithm}. We make the important distinction that our work focuses on \textbf{filtering} the best test cases to fit into a prompt, \textbf{not generating} said test cases from scratch, because large quantities of test cases are often readily available in real-world applications~\cite{lemieux2023codamosa}, and generation from scratch entails ensuring input-output pairs both describe ground truth code and maximize Pass@1. Nonetheless, our filtered examples, $\tilde{D}_j$, would benefit from better pool of demonstration examples, $\hat{D}_j$, from improved generation algorithms.

\begin{table*}
    \centering
    \begin{tabular}{c|l|l|l|l}
         LLM &  No Example& Best Example&Average Example& Median Example\\
         \hline
         \tfsmall &  $(6.62 \pm 1.68) \cdot 10^8$ & $(1.38 \pm 0.41)\cdot 10^6$ & $(3.09 \pm 1.01) \cdot 10^8$ & $(2.77 \pm 1.08) \cdot 10^8$\\
         \tfbase &  $(4.09 \pm 2.37) \cdot 10^6$ & $(1.62 \pm 0.99) \cdot 10^6$&$(4.34 \pm 1.49) \cdot 10^6$ & $(3.72 \pm 1.48) \cdot 10^6$\\
         \mixtral &  $4.73 \pm 0.26$& $2.31 \pm 0.09$ & $4.67 \pm 0.27$ & $4.33 \pm 0.22$\\
         \llama &  $3.20 \pm 0.27$& $2.15 \pm 0.09$ & $2.80 \pm 0.14$ & $2.65 \pm 0.13$\\
         \codellama & $3.21 \pm 0.11$&$1.76 \pm 0.04$ & $3.35 \pm 0.15$ &$3.15 \pm 0.12$\\
    \end{tabular}
        \vspace{-0.5cm}

    \caption{Studying the performance (i.e. $PP_{target}(y_m^{(j)}; f_m, [(x_c^{(j,i)}, y_c^{(j,i)})])$) by adding a single few-shot example to each prompt. We report the average and best performance gain from adding single examples. Standard error is computed across different prompts.}
    \label{tab:msize}
\end{table*}
\section{Methodology}
\label{sec:method}

The goal of our work is to find demonstration examples, $\tilde{D}_j$, that outperforms the human defined prompts in \hevp~\citep{liu2024your}, $D_j$. We accomplish this by proposing 2 algorithms for $f_p$ that approximates the Pass@1 metric when an example, $[(x_c^{(j,i)}, y_c^{(j,i)})]$, is added to the prompt: $\mathcal{T}(u_{NL}^{(j)},[(x_c^{(j,i)}, y_c^{(j,i)})])$. Because Pass@1 is a discrete metric, we optimize a more granular surrogate objective: the perplexity of generating the ground truth code, \perptgt, as described in Equation~\ref{eqn:pcp_tgt}, where perplexity is defined as $PP(y; f_m, x) = Pr_{f_m} \left [ y | x \right ] ^ {-\frac{1}{|y|}}$~\citep{jelinek1977perplexity}. We show that \perptgt's ranking is a good estimator for Pass@1's ranking in Section~\ref{sec:pcp_exp}. 


\begin{multline}
PP_{target}(y_m^{(j)}; f_m, (x_c^{(j,i)}, y_c^{(j,i)})) = \\ PP \left (y_m^{(j)} ; f_m, \mathcal{T}(u_{NL}^{(j)},[(x_c^{(j,i)}, y_c^{(j,i)})]) \right )
\label{eqn:pcp_tgt}
\end{multline}


\begin{figure}
    \centering
    \includegraphics[width=1.0\linewidth]{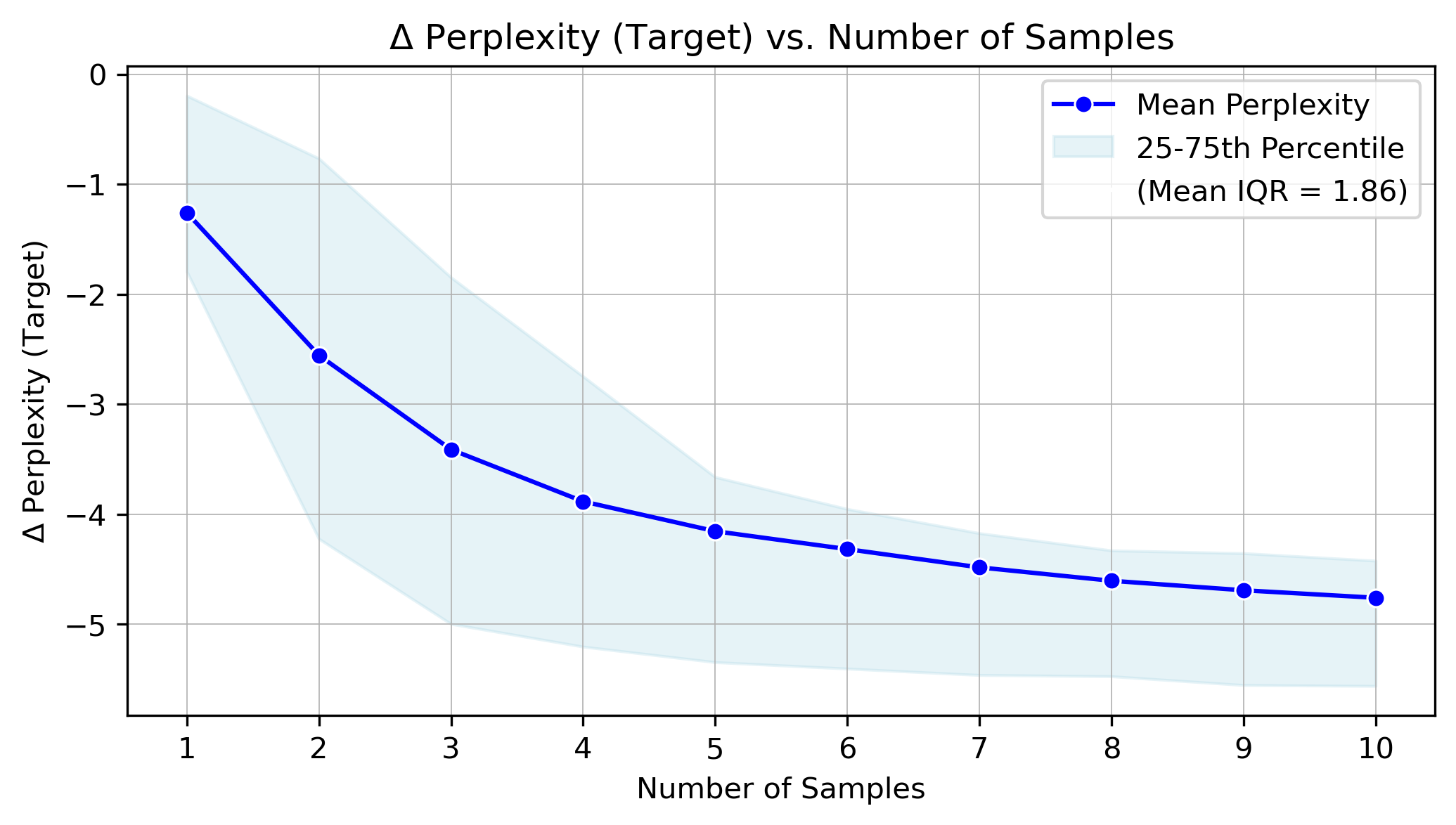}
    
    \caption{Studying the performance gain (i.e. $PP_{target}(y_m^{(j)}; f_m, \hat{D}_j)-PP_{target}(y_m^{(j)}; f_m, \emptyset)$) from adding multiple examples, $\tilde{D}_j$ to each prompt. Each example is chosen randomly from a larger pool of candidate examples, $\hat{D}_j$. Deviations are computed across different prompts.}
    \label{fig:exp_more}
\end{figure}

\subsection{Model-Free Ranker}
\label{sec:method-model-free}

As mentioned in Section~\ref{sec:method}, the goal of $f_p$ is to approximately rank the Pass@1 metric by ranking the \perptgt metric. Inspired by recent work showing foundation models learn more from data they cannot generate~\citep{shumailov2024ai}, our model-free ranker makes the following qualitative assumptions: (1) all few-shot examples will improve performance by disambiguating the natural language prompt, and (2) few-shot examples that the LLM cannot generate are more effective at disambiguating the natural language prompt. We show the degree to which these assumptions hold in Section~\ref{sec:whether}.

Under these assumptions, the model-free algorithm selects few-shot examples that the LLM cannot self-generate, which we measure with \perpsrc as described in Equation~\ref{eqn:pcp_src}.

\begin{multline}
PP_{source}((x_c^{(j,i)}, y_c^{(j,i)}); f_m) = \\ PP \left ( \mathcal{T}(u_{NL}^{(j)},[(x_c^{(j,i)}, y_c^{(j,i)})]); f_m, \emptyset \right )
\label{eqn:pcp_src}
\end{multline}

We show that \perpsrc's ranking is a good estimator for \perptgt's ranking in Section~\ref{sec:pcp_exp2}. Intuitively, if the LLM does not expect to read a certain few-shot example, then this example will have a large effect on the LLM's generated output. As long as this effect is positive (the model does not hallucinate), \perpsrc remains a reliable metric for selecting impactful few-shot examples. \oursfree, $f_p^{(free)}$, ranks the examples based on descending \perpsrc.




\subsection{Model-Based Ranker}
\label{sec:method-model-based}

To generate rankings with less assumptions, we introduce a model-based algorithm for $f_p$. Our model-based algorithm determines whether examples improve or deteriorate performance by directly learning trends from a labelled training dataset. Specifically, we collect a dataset of few-shot example, $[(x_c^{(j,i)}, y_c^{(j,i)})]$, to \perptgt pairs and train a neural network, $f_b$, to predict the \perptgt based on LLM representations of the labelled example: $f_b(f_m(\cdot); \theta)$. \oursmodel, $f_p^{(based)}$ ranks the examples based on descending outputs of the neural network.

\textbf{Collecting the Dataset:} We collect the training dataset by splitting the \hevp dataset into the training and testing splits in a 80:20 split, ensuring prompts in the test set do not leak into the training set. For each training set prompt, we form new labelled pairs from each few-shot example, where inputs are $x_b^{(j,i)} = (u_{NL}^{(j)},[(x_c^{(j,i)}, y_c^{(j,i)})])$ and outputs are $y_b^{(j,i)} = PP_{target}(y_m^{(j)}; f_m, (x_c^{(j,i)}, y_c^{(j,i)}))$.


\textbf{Extracting Embeddings:} We pass the input prompt to \codellama~\citep{roziere2023code} to obtain hidden representations. Because intermediate layers of LLMs contain more semantic information, we extract the [EOS] token from the $16$th layer, which we denote as $h_b^{(j,i)} = f_m(\mathcal{T}(u_{NL}^{(j)},[(x_c^{(j,i)}, y_c^{(j,i)})]))$. This embedding contains rich semantic information on each test case, as shown in Section~\ref{sec:pcp_exp3}.

\textbf{Model:} We use a simple 4-layer multi-layer perceptron (MLP) model, $f_b$, to decode the \codellama embedding. We train the MLP using mean squared error on the training dataset with the LLM being frozen: $MSE(f_b(h_b^{(j,i)}), y_b^{(j,i)})$. Unlike \oursfree, \oursmodel is directly trained on the \perptgt metric, hence can better distinguish cases where when the example hurts or improves the LLM's coding capabilities.




\begin{figure*}
    \centering
    \includegraphics[width=1.0\linewidth]{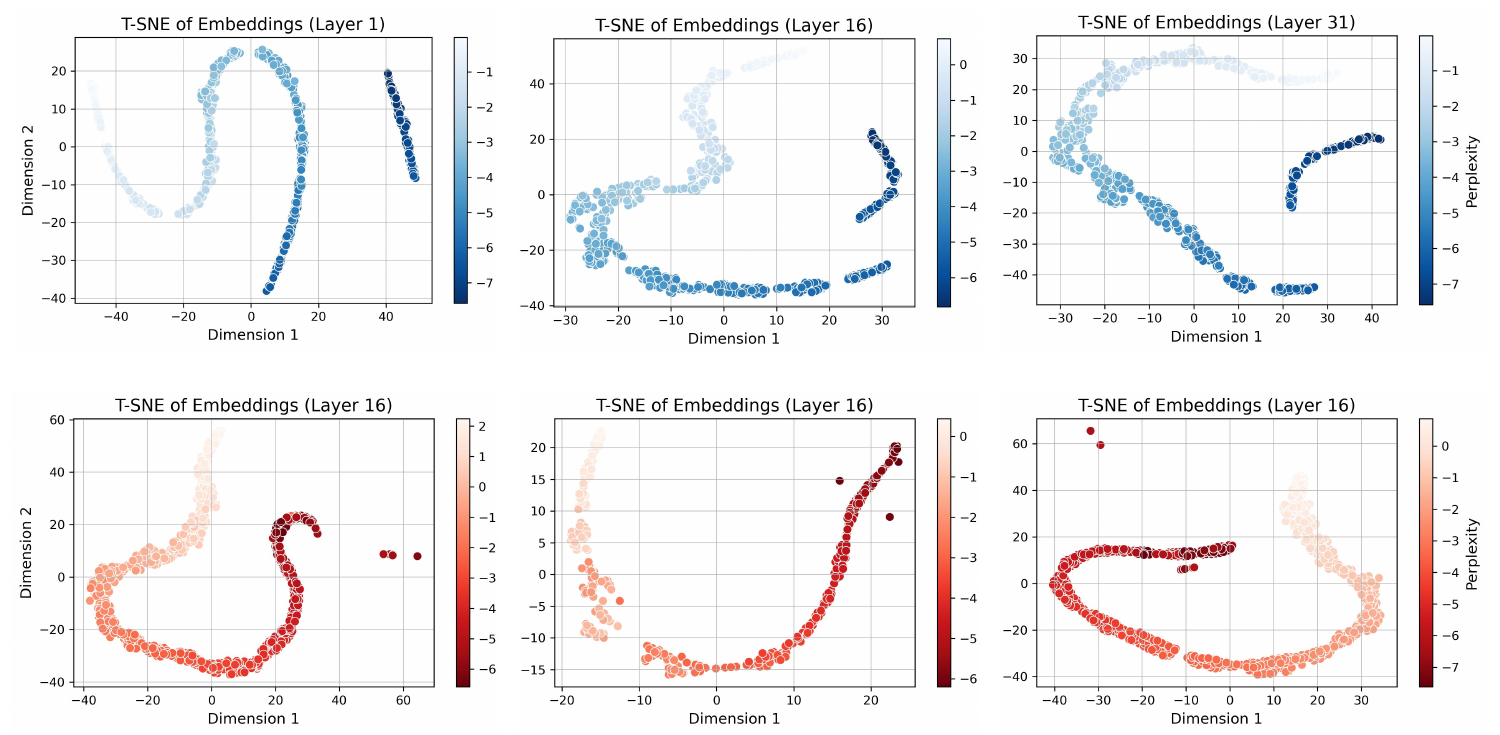}
    \caption{t-SNE Plots of the hidden representation of the prompt with given examples, $h_b^{(j,i)}$. Colors denote the change in \perptgt. As shown, the embeddings naturally encode the \perptgt score.}
    \label{fig:tsne}
\end{figure*}

\begin{figure}
    \centering
    \includegraphics[width=1.0\linewidth]{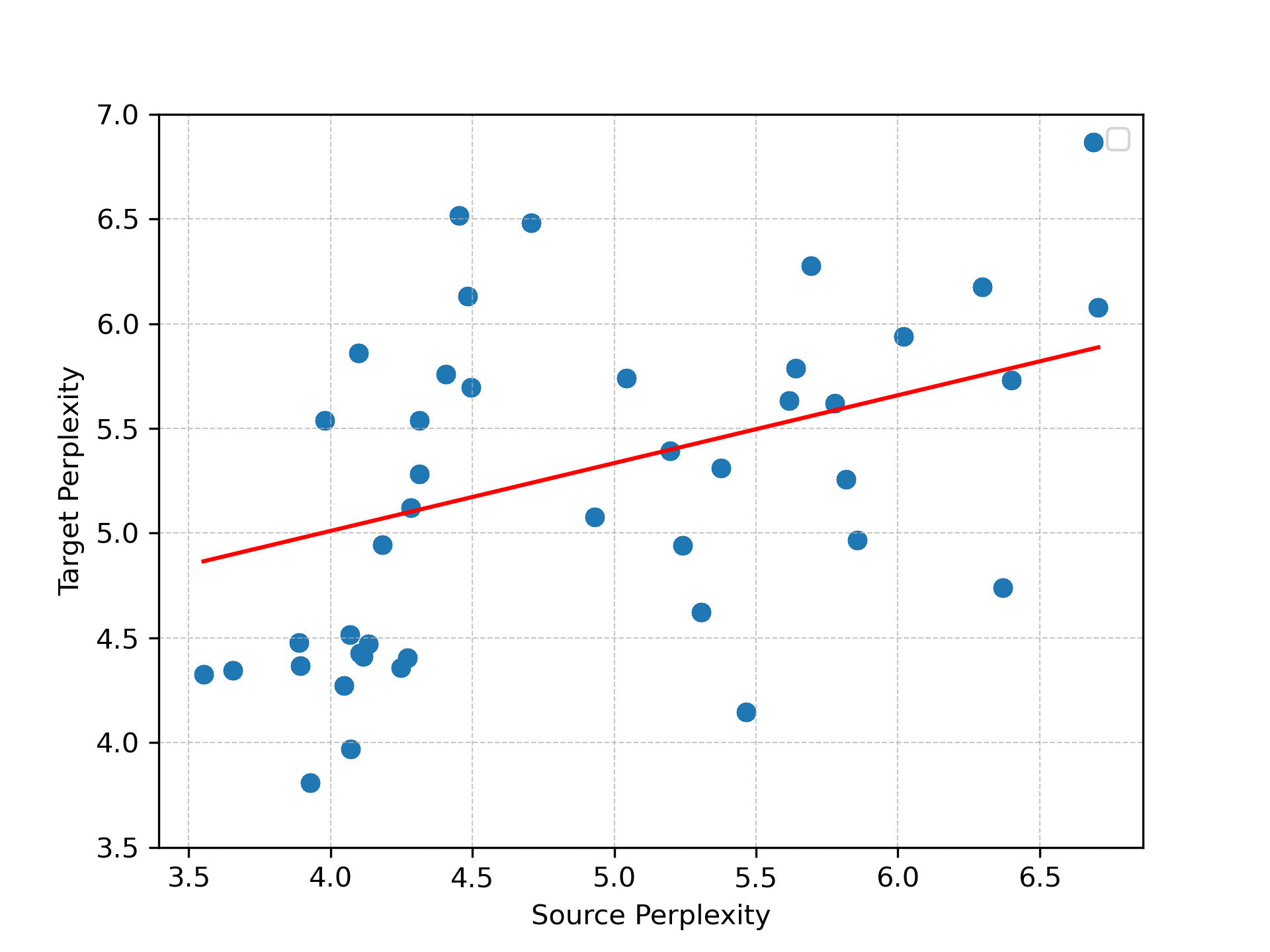}
    \caption{Correlation between \perpsrc and \perptgt on some randomly chosen examples across some randomly chosen problems.}
    \label{fig:src_interp}
\end{figure}

\begin{table}
    \centering
    \begin{tabular}{c|c|c}
         &  Pass@1=0.0& Pass@1=1.0\\
         \hline
         \textsc{Perplexity} & \multirow{2}{*}{$0.21\pm0.19$} & \multirow{2}{*}  {$0.64\pm0.15$}\\
         \textsc{(target)} &  & \\
    \end{tabular}
    \caption{Average \perptgt of test cases that have Pass@1 score of 1.0 and of test cases that have Pass@1 score of 0.0. Results are aggregated across all coding problems. Deviation reported is standard error.}
    \label{tab:tgt_interp}
\end{table}



\begin{figure}
    \centering
    \includegraphics[width=1.0\linewidth]{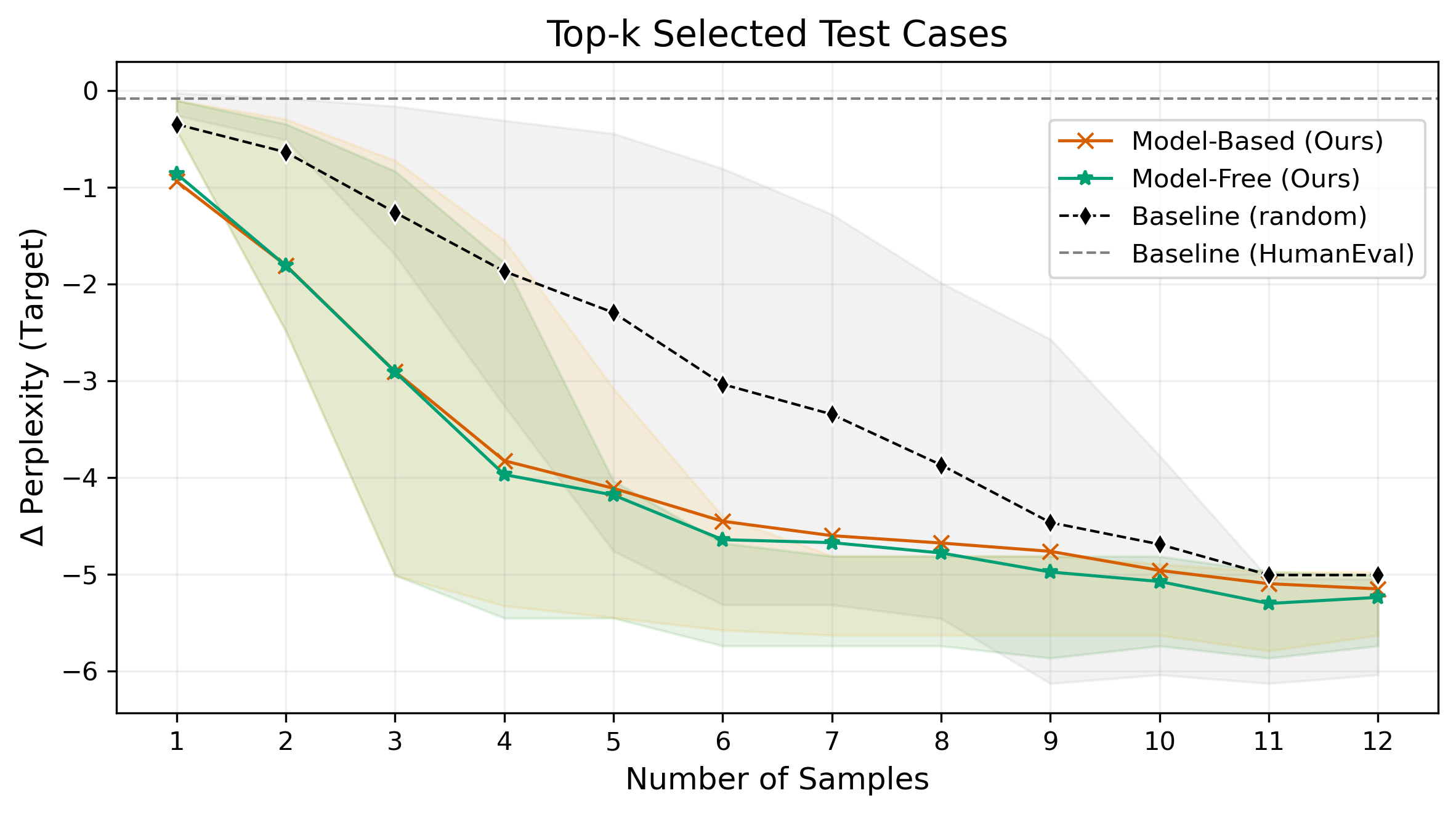}
    \caption{Main Results using \perptgt. We compare the perplexity improvement gained from adding a $N$ examples to the prompt, where each example is chosen by a different $f_p$ ranker function.}
    \label{fig:perplexity}
\end{figure}

\begin{figure*}
    \centering
    \includegraphics[width=1.0\linewidth]{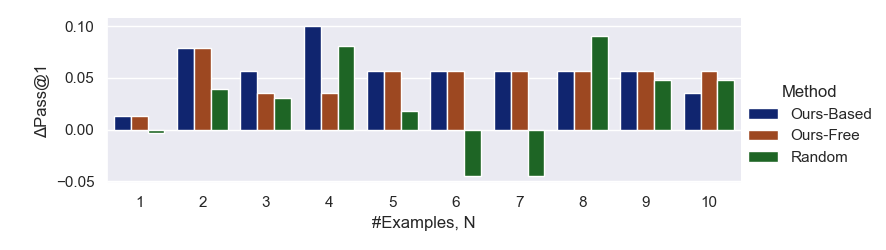}
    \caption{Main Results using Pass@1. We compare the perplexity improvement gained from adding a $N$ examples to the prompt, where each example is chosen by a different $f_p$ ranker function.}
    \label{fig:passat1}
\end{figure*}

\begin{figure}
\centering
\begin{subfigure}{0.4\textwidth}
  \centering
  \includegraphics[width=1.0\linewidth]{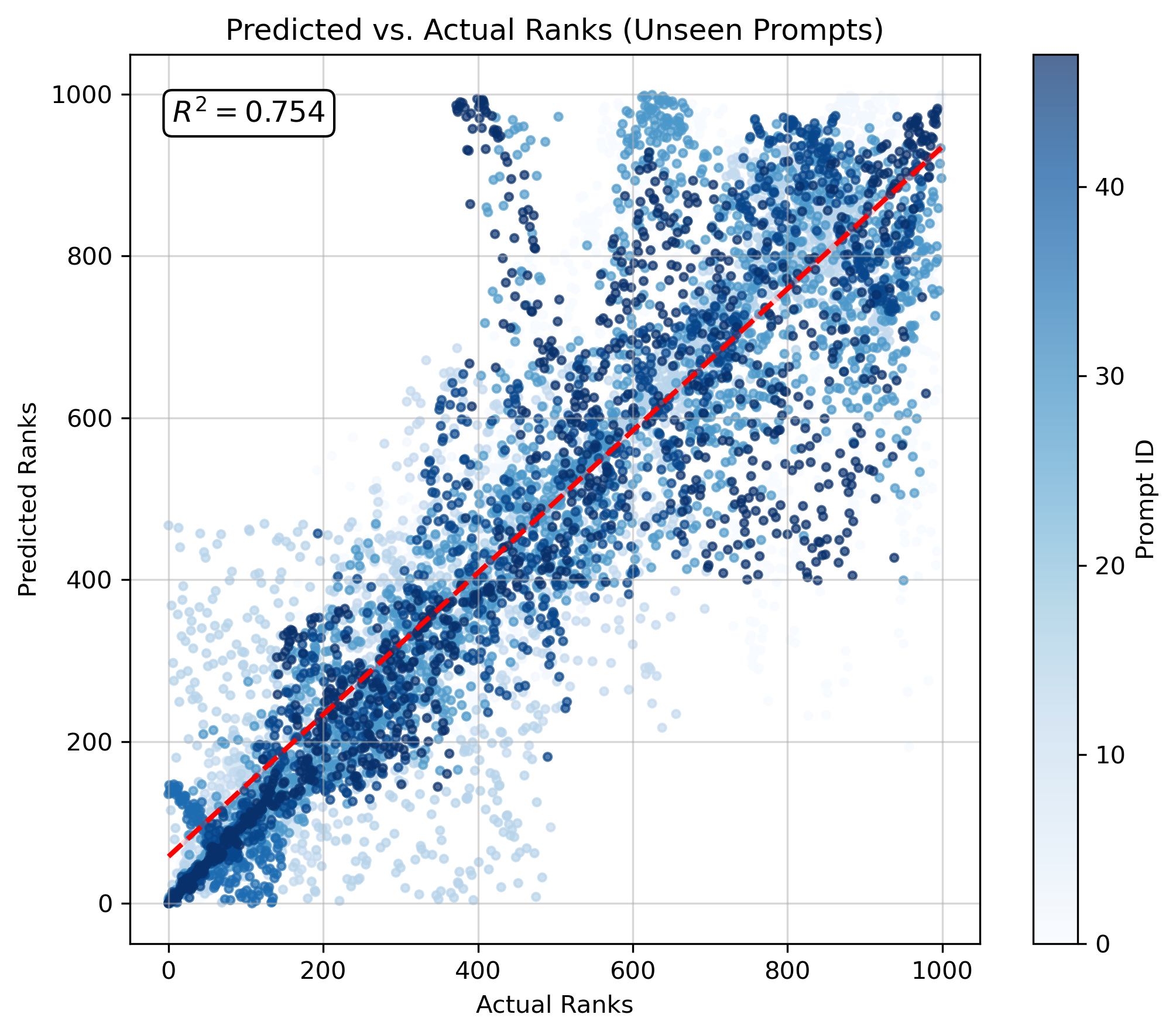}
  \caption{Out-of-Distribution}
  \label{fig:sub1}
\end{subfigure}%
\\
\begin{subfigure}{0.4\textwidth}
  \centering
  \includegraphics[width=1.0\linewidth]{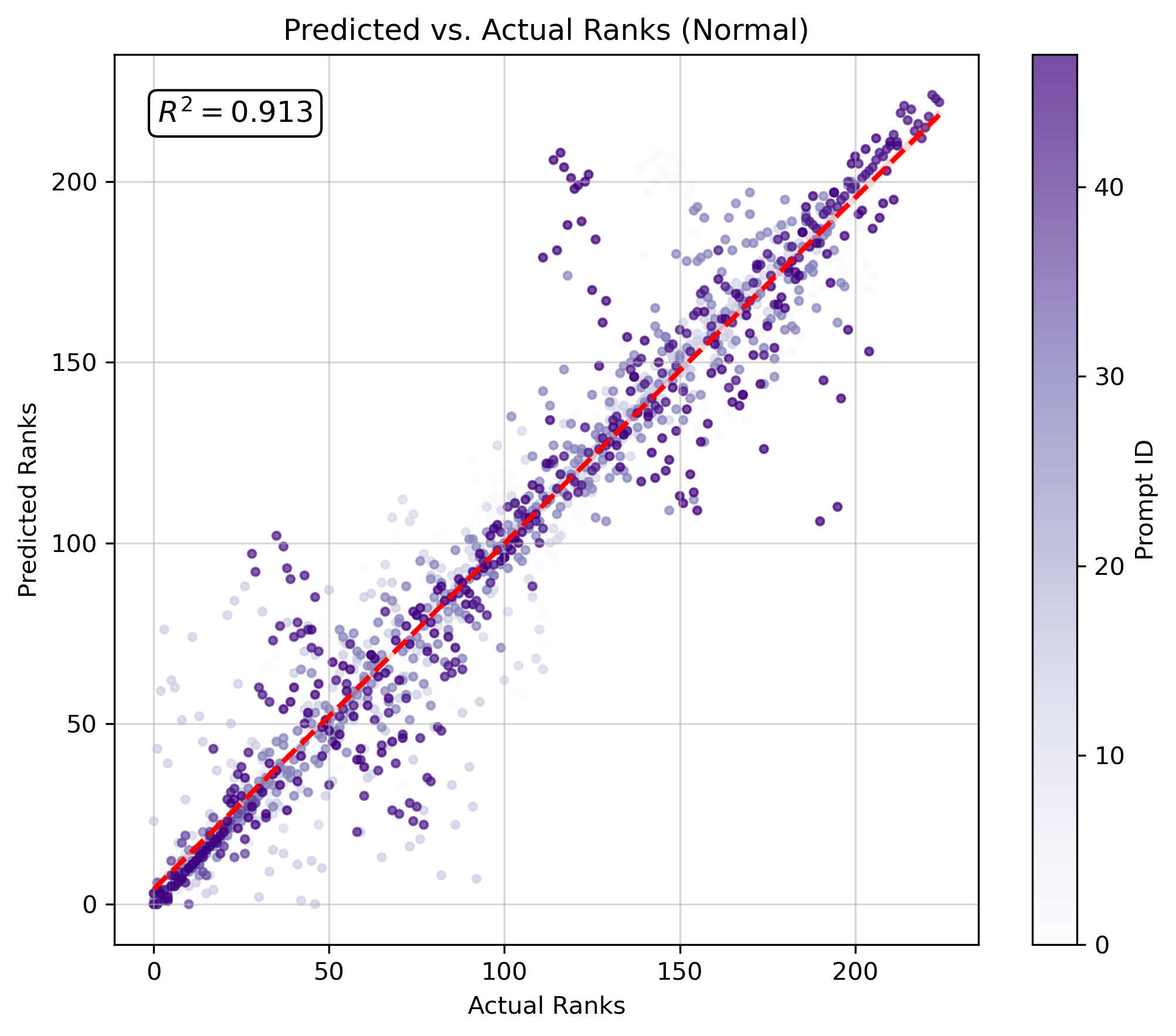}
  \caption{In-Distribution}
  \label{fig:sub2}
\end{subfigure}
\caption{Measuring the effect in-distribution and out-of-distribution training has on the model's performance. All other experiments in this work adopt the harder out-of-distribution training, where the prompts are split in to train test sets that are never seen in training. In contrast, the in-distribution training splits the examples within a prompt. Hence, the sample natural language prompt can appear in both training and test sets. The ``Actual Ranks'' is the \perptgt ranking and the ``Predicted Ranks'' is the \oursmodel ranking.}
\label{fig:spearman}
\end{figure}


\section{Motivational Experiments}
\label{sec:motiv_exp}

\subsection{Models, Datasets, and Metrics}
\label{sec:exp_prelim_split}

We evaluate the effect of adding few-shot examples on 5 language models: \tfsmall (60M)~\citep{raffel2020exploring}, \tfbase (220M)~\citep{raffel2020exploring}, \mixtral (7B)~\citep{jiang2023mistral}, \llama (8B)~\citep{dubey2024llama}, and \codellama (7B)~\citep{roziere2023code}. We evaluate \oursfree ($f_p^{(free)}$) and \oursmodel ($f_p^{(based)}$) on the best LLM among the above choices, \codellama ($f_m$) on the popular \hevp benchmark~\citep{liu2024your}. As described in Section~\ref{sec:method-model-based}, we divide the prompts into 2 sets, one for training \oursmodel and one for validation/testing in an 60:20:20 split. In this work, we adopt GPT~\citep{radford2019language} to generate new few-shot examples for each test prompt. However, because the \hevp benchmark~\citep{liu2024your} already uses GPT to generate its test cases, to prevent data leakage, the large pool of examples is obtained from \hevp benchmark. Specifically, we split the unit tests into (1) few-shot examples used for prompt optimizations ($\hat{D}_j = [(x_c^{(j,i)}, y_c^{(j,i)})]_{i=0}^{M}$) and (2) few-show examples used for benchmarking in an 60:20:20 split. 

We note that the Pass@1 metric is too coarse for our use-case. Specifically, state-of-the-art models~\citep{guo2024deepseek,huang2023agentcoder} show minimal performance differences between different algorithms. To better measure the effect of prompt design, we use the more finegrained \perptgt metric, described in Section~\ref{sec:method}. We show the correlation between Pass@1 and \perptgt in Section~\ref{sec:pcp_exp}. We show that few-shot example selection improves Pass@1 results in Section~\ref{sec:main_exp}.


\subsection{Whether Few-Shot Examples Help}
\label{sec:whether}

We first evaluate \textbf{whether} few-shot example help or hurt downstream performance. First, we test whether adding \textbf{single} few-shot examples improves the LLM's probability of generating the ground truth solution, \perptgt, by iterating through \textit{all} possible single test cases, $PP_{target}(y_m^{(j)}; f_m, [(x_c^{(i,j)},y_c^{(i,j)})]) \forall i \forall j$. Next, we test whether adding \textbf{multiple} few-shot examples improves the LLM's probability of generating the ground truth solution, $\Delta$\perptgt, formally defined as $PP_{target}(y_m^{(j)}; f_m, \hat{D}_j)-PP_{target}(y_m^{(j)}; f_m, \emptyset)$.


Our study shows prompt design has a large influence on the downstream performance. Specifically, \textbf{by adding just a single one-shot example, LLM coding capabilities can improve by the same margin as model architecture or training dataset design choices (Table~\ref{tab:msize})}. Following neural scaling laws~\citep{kaplan2020scaling}, larger models, such as \mixtral, \llama, and \codellama, exhibit much better zero-shot coding capabilities than \tfsmall and \tfbase. We find \textbf{most examples improve \perptgt}, because the median test case improves \perptgt. However, not all test cases are created equal. \textbf{The best test case for a given problem improves \perptgt much more than the median test case}, motivating a need to rank candidate examples, $f_p$, by their helpfulness. Among the LLMs tested, \codellama~\citep{roziere2023code} benefits the most from good test case selection, hence we focus this work around \codellama.

Given the improvement from single few-shot examples, we extend our analysis to multiple test cases. By adding more few-shot examples, \textbf{\codellama coding performance significantly improves then saturates} around a log perplexity improvement of around 5.0 after 6 examples (Figure~\ref{fig:exp_more}). While performance monotonically improves, longer prompts require much more compute costs~\citep{dao2022flashattention}. Specifically, \textbf{the LLM is approximately $2\times$ slower on prompts including 10 few-shot examples compared to prompts with no examples (only the natural language).} Thus, there is a natural tradeoff between performance and compute costs in terms of the number of examples.





Because \textbf{both single and multiple few-shot examples improves LLM's coding performance}, we showed prompt optimization on its own can improve the LLM's code generation abilities without any additional LLM training, model design, or model alignment. As mentioned previously, LLM comput costs are determined by the prompt token length, motivating a need to select the best $N\ll M$ few-shot examples from the larger pool of $M$ many-shot examples.



\subsection{Which Few-Shot Examples Help}

Because most few-shot examples improve \codellama coding capabilities, we shift our focus to studying \textbf{which} examples improve its coding capabilities the most. 



\subsubsection{Target Perplexity}
\label{sec:pcp_exp}

As explained in Section~\ref{sec:method}, Pass@1 is a discrete metric, hence cannot capture the more fine-grained performance gains from different design choices. Specifically, \textbf{the Pass@1 metric is binary on a single coding problem}. Thus, Pass@1 does not effectively measure the effect of including individual one-shot examples in the prompt of a single problem. Hence, we propose a more finegrained surrogate metric, \perptgt, to measure code performance. \textbf{Our analysis (Table~\ref{tab:tgt_interp}) confirms Pass@1 scores correlate with the \perptgt scores, since successful prompts have higher mean perplexity.} Unlike Pass@1, \perptgt is a more granular evaluation metric that measures the effect of few-shot examples on individual problems.




\subsubsection{Source Perplexity}
\label{sec:pcp_exp2}

\oursfree estimates the \perptgt ranking by few-shot examples by their \perpsrc scores. Following the one-shot experimental setup from Section~\ref{sec:whether}, \textbf{our analysis verifies \perpsrc somewhat approximates the \perptgt scores (Figure~\ref{fig:src_interp}).} We highlight the benefit of \perpsrc is it does \textit{not} require access to any ground truth code to compute. Hence this metric can directly be used for model-free inference, as elaborated in Section~\ref{sec:method-model-free}.



\subsubsection{Embedding-Level Signals}
\label{sec:pcp_exp3}

While \perpsrc roughly correlates ~\perptgt, we discover LLM embeddings themselves also provides a reliable signal on a prompt's \perptgt score. Specifically, we plot the t-SNE 2D visualization of the 16th layer of \codellama. \textbf{Our investigation (Figure~\ref{fig:tsne}) reveals LLM embeddings themselves are a suitable signal to measure \perptgt.} These results motivate for our learning-based approach, \oursmodel, which encodes semantic representations of one-shot examples to estimate \perptgt. 

\begin{table*}
    \centering
    \begin{tabular}{c| c| c}
         & Prompt: Find the factors & Prompt: Find the largest divisor \\
         \hline
         Best Example& factorize(1000) == [2, 2, 2, 5, 5, 5]& largestDivisor(42) == 21\\
         2nd Best Example& factorize(2147483647) == [2147483647]& largestDivisor(55) == 11\\
         3rd Best Example& factorize(1024) == [2, 2, 2, 2, 2, 2, 2, 2, 2, 2]& largestDivisor(44) == 22\\
         ...&  ...& ...\\
         3rd Worst Example& factorize(79) == [79] & largestDivisor(79) == 1\\
         2nd Worst Example& factorize(67) == [67]& largestDivisor(53) == 1\\
         Worst Example& factorize(83) == [83]& largestDivisor(71) == 1 \\
    \end{tabular}
    \caption{Case Studies showing which one-shot examples most improve \perptgt.}
    \label{tab:casestudy}
\end{table*}

\subsection{Case Study Analysis}

To understand which examples boost LLMs coding capabilties, we plot the top 3 most informative and least informative prompts for 2 random prompts. We observe that\textbf{ prompts with more complex inputs tend to be informative (Table~\ref{tab:casestudy}), while simple or edge case examples provide little benefit.} As shown in Section~\ref{sec:main_exp}, \oursfree and \oursmodel boost model performance by selecting these informative few-shot examples quantitatively and automatically.


\section{Main Experiments}
\label{sec:main_exp}
Given the motivating results in Section~\ref{sec:motiv_exp}, we now evaluate whether our proposed \textbf{ranking function,} $f_p$ improves model performance. We adopt the same dataset splitting strategy as in Section~\ref{sec:exp_prelim_split}.

\subsection{Main Results}


Our experiments show \textbf{both \oursfree and \oursmodel improves \codellama's likelihood of generating ground truth code by up to 5.0 through solely optimizing the prompt (Figure~\ref{fig:perplexity}).} Specifically, under the fixed $N$-shot setting, \oursfree extracts more effective prompts than both the original prompt (Human Eval), randomly selecting few-shot examples (Random). Figure~\ref{fig:perplexity} also demonstrates our methods improve perplexity over the natural language prompt on its own.


Not only does our approach improve the \perptgt score, but \textbf{both \oursfree and \oursmodel improves \codellama's raw Pass@1 coding performance by $5.70\pm2.17\%$ and $5.05\pm1.70\%$ on average respectively through solely optimizing the prompt (Figure~\ref{fig:passat1}).} We highlight \oursmodel performs better than the random choice under all settings of $N$. These results indicate that our prompt optimization techniques have practical application.




\subsection{Distribution Shift Bottleneck}

We provide further analysis into future directions for \oursmodel. Specifically, we find \oursmodel can be improved by a better training dataset. We arrive at this conclusion by splitting across all few-shot examples instead of splitting across all prompts. Under this setting, the same prompt can occur between training and testing. Hence, the model only needs to learn the importance of individual examples, rather than learn the importance of individual examples and generalize to new prompts. By removing the need to generalize to new pormpts, \oursmodel greatly improves in its ranking predictions (Figure~\ref{fig:spearman}). This indicates better dataset design which improves the prompt generalization of \oursmodel can further improve its performance, encouraging development of larger coding datasets. We highlight all the experiments in this work adopt the harder setting, where the train and test set have disjunct natural language descriptions .

\section{Conclusion}

In this work, we analyzed the effects of few-shot examples on the coding capabilities of large language models (LLMs). Our work identified interesting properties of few-shot examples, including the example complexity, embedding representation, and perplexity. To this end, we proposed 2 effective strategies of picking in-context examples, a model-free algorithm based on perplexity, \oursfree, and a model-based algorithm trained on data, \oursmodel. We show that both \oursfree and \oursmodel meaningfully improves \codellama's coding capabilties.

\section{Limitations}

This work studies several aspects of few-shot examples in the prompt. The main experiments are performed on \codellama~\citep{roziere2023code}. For future work, we wish to extend to more state-of-the-art models such as DeepSeeker~\citep{guo2024deepseek}. This work does not tackle chain of thought prompting, only the original prompt in a zero-shot manner. This work focuses on code generation. While the techniques are applicable to general in-context learning, we leave such study to future work. This work tests on the \hevp~\citep{liu2024your} benchmark. We leave incorporating harder benchmarks, such as BigCodeBech~\citep{zhuo2024bigcodebench}, to future work.

\bibliography{custom}

\appendix

\section{Distribution Plots across Language Models}
\begin{figure}
    \centering
    \includegraphics[width=1.0\linewidth]{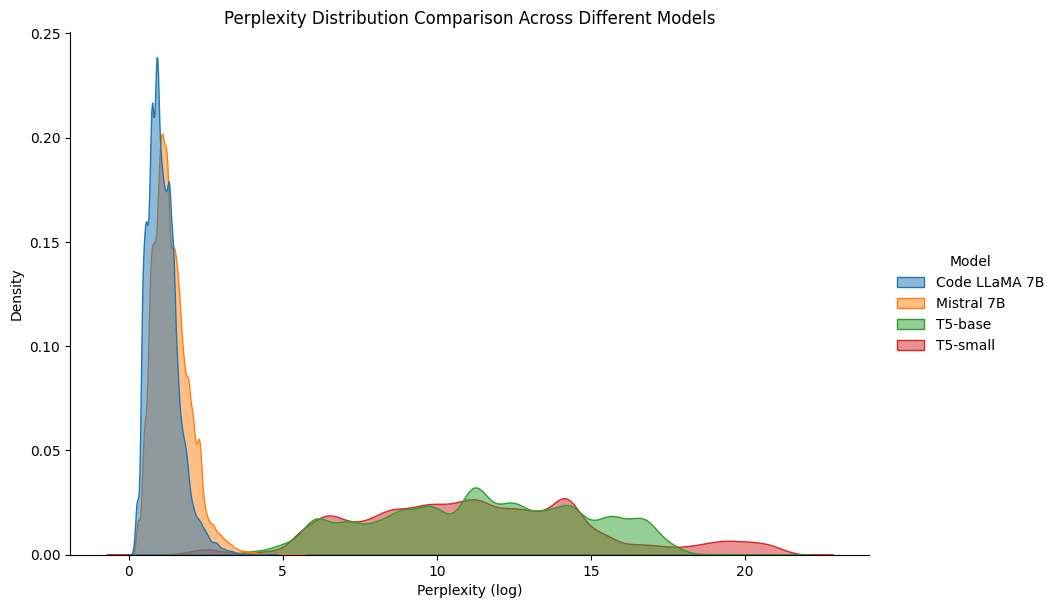}
    \caption{Comparing the \perptgt distribution between \tfbase, \tfsmall, \mixtral, and \codellama. The x axis is the \perptgt in log scale.}
    \label{fig:comp_perp_between_models}
\end{figure}

Figure \ref{fig:comp_perp_between_models} compares the \perptgt between different models. As expected the larger models (\mixtral and \codellama) have lower perplexity generating the test prompts compared to the smaller models (\tfbase and \tfsmall). Furthermore \codellama is more specialized in coding tasks thus yielding slightly lower \perptgt score compared to \mixtral despite their comparable sizes. 

\section{Hyperparameter Settings}
\label{sec:hyperparameter}

Because \oursfree is a model-free algorithm with no hyperparameters. For \oursmodel, we use the Adam optimizer with a learning rate of 1e-4 for 1000 epochs. \oursmodel uses a 4-layer neural network with hidden dimension sizes 256, 128, 64, and 1, batch normalization, and ReLU activation. We also considered 2- and 3-layer neural networks. \oursmodel uses the 16th layer's EOS token. We also considered the average token embedding or BOS token embedding from any other layer. We chose our settings by evaluating on the validation set.

\section{Machine and Compute Times}

All experiments were conducted across 8 Nvidia V100 GPU and an AMD EPYC 7402 CPU. \oursfree and the random baseline takes around 15 minutes to run through the whole benchmark of 7000 test cases. \oursmodel trains in under 5 minutes and takes around 15 minutes to run through the whole benchmark of 7000 test cases. Collecting the training dataset, specifically the prompt embedding takes the longest time, at around 11 hours. Our work is built using pyTorch~\citep{paszke2019pytorch}, Transformers~\citep{wolf-etal-2020-transformers}, Numpy~\citep{harris2020array}, Pandas~\citep{mckinney-proc-scipy-2010}, Matplotlib~\citep{Hunter:2007}, and Seaborn~\citep{Waskom2021}.

\section{Dataset}

All experiments are conducted on the popular \hevp~\citep{liu2024your} benchmark. This dataset augments the original HumanEval~\citep{chen2021evaluating} dataset with GPT-generated test cases and further test cases from mutation genetic algorithms. We split the test cases into evaluation test cases and those used for few-shot examples in an 80:20 split. We split the prompts into training and testing in an 80:20 split. We follow the intended usage of the datasets.

\section{Large Language Models}

We considered several large language models in this work. Our experiments are primarily performed on \codellama~\citep{roziere2023code}, which is a 7B decoder-only model trained on code synthesis. We also explore \tfsmall and \tfbase~\citep{raffel2020exploring}, which are 60M and 220M encoder-decoder models trained on many different tasks, \mixtral~\citep{jiang2023mistral}, which is an 7B decoder-only model trained for language generation, and \llama~\citep{dubey2024llama}, which is a 8B decoder-only model also trained for language generation. We follow the intended usage of the LLMs.








\end{document}